\documentclass[journal=jctc,manuscript=article]{achemso}

\usepackage[version=3]{mhchem} 
\usepackage[T1]{fontenc}       

\usepackage{caption} 
\usepackage{xspace} 
\usepackage{varwidth}
\usepackage[titletoc]{appendix}

\usepackage{graphicx}
\usepackage{lscape}
\usepackage{float}
\usepackage{psfrag}
\usepackage{amssymb}

\SectionNumbersOn

\author{Simon P. Neville}
\affiliation{National Research Council of Canada, 100 Sussex
  Drive, Ottawa, Ontario K1A 0R6, Canada}
\email{Simon.Neville@nrc-cnrc.gc.ca}

\author{Michael S. Schuurman}
\affiliation{National Research Council of Canada, 100 Sussex
  Drive, Ottawa, Ontario K1A 0R6, Canada}
\alsoaffiliation{Department of Chemistry and Biomolecular Sciences,
  University of Ottawa, 10 Marie Curie, Ottawa, Ontario, K1N 6N5,
  Canada}
\email{Michael.Schuurman@nrc-cnrc.gc.ca}

\title{Removing the deadwood from DFT/MRCI wave functions: The
  p-DFT/MRCI method}

\begin{document}

\begin{abstract}
  The combined density functional theory and multireference
  configuration interaction (DFT/MRCI) method is a powerful tool for
  the calculation of excited electronic states of large
  molecules. There exists, however, a large amount of superfluous
  configurations in a typical DFT/MRCI wave function. We show that
  this deadwood may be effectively removed using a simple
  configuration pruning algorithm based on second-order Epstein-Nesbet
  perturbation theory. The resulting method, which we denote
  p-DFT/MRCI, is shown to result in orders of magnitude saving in
  computational timings, while retaining the accuracy of the original
  DFT/MRCI method.
\end{abstract}

\section{Introduction}
Since its introduction by Grimme and Waletzke\cite{grimme_dft-mrci},
the combined density functional theory and multireference
configuration interaction (DFT/MRCI) method has proved to be a
uniquely powerful tool for the calculation of the excited states of
large molecular systems. The method was originally conceived as a
means to calculate singlet and triplet valence excited states of
organic molecules. However, the scope of the method has recently been
significantly widened to include multiplicity-independent
formulations\cite{lyskov_dftmrci_redesign, heil_dftmrci_2017}, a
parameterization tuned to the description of transition metal
complexes\cite{heil_dftmrci_transition_metals}, the treatment of
core-excited states\cite{seidu_cvsdftmrci}, and the calculation of
spin-orbit and non-adiabatic
couplings\cite{kleinschmidt_spin_orbit_coupling_constants,
  kelinschmidt_spock_ci} and diabatic
potentials\cite{ours_dftmrci_pbdd}.

At its heart, DFT/MRCI is an individually selecting MRCI method,
utilizing short MRCI expansions to describe static electronic
correlation and DFT-specific Hamiltonian matrix corrections to capture
the remaining dynamic correlation\cite{marian_dftmrci_review}. As a
result, DFT/MRCI wave function expansions are extremely compact
relative to those in canonical MRCI methods. Experience has shown,
however, that DFT/MRCI wave functions typically contain a large amount
($\gtrsim$90\%) of deadwood configurations. That is, configurations
that contribute negligibly to the norm of a DFT/MRCI wave function. If
these configurations could be identified \textit{a priori} and
removed, then large computational savings would be afforded.

The study of algorithms for the removal of superfluous configurations
from CI calculations dates back to the earliest days of the field with
the pioneering work of Davidson\cite{davidson_selected_ci_1969} and
Whitten and Hackmeyer\cite{whitten_selected_ci_1969,
  hackmeyer_selected_ci_1971}. Other notable early examples include
the multi-reference double-excitation CI (MRD-CI) method of Buenker
and Peyerimhoff\cite{beunker_mrdci_1974, beunker_mrdci_1975,
  buenker_mrdci} and the CI by perturbation with multiconfigurational
zeroth-order wavefunction selected by iterative process (CIPSI) method
of Huron \textit{et al.}\cite{huron_cipsi_1973}. Recent years have
seen a resurgence of interest in such selected CI methods. Prominent
examples include the refinement of the CIPSI methodology by Scemama
and co-workers\cite{scemama_quantum_package, scemama_cipsi_benzene,
  garniron_thesis}, the semistochastic heat-bath CI
(SHCI)\cite{umrigar_2016, umrigar_2017, sharma_2017, umrigar_2018},
adaptive sampling CI (ASCI)\cite{tubman_2016, tubman_2020}, adaptive
CI (ACI)\cite{schriber2016, schriber2017}, iterative CI
(iCI)\cite{hoffmann_2020, hoffmann_2021}, and iterative configuration
expansion (ICE)\cite{neese_2021_I, neese_2021_II} methods. Common to
many of the above mentioned selected CI methods is the use of aspects
of second-order Epstein-Nesbet perturbation theory (ENPT2) to estimate
the importance of individual determinants or configuration state
functions (CSFs) via their interaction with a set of zeroth-order
eigenfunctions.

The use of ENPT2 to estimate \textit{a priori} the importance of
individual configurations is readily applicable within the framework
of the DFT/MRCI method, and it is the purpose of this paper to explore
this as a route to the elimination of the deadwood from DFT/MRCI wave
functions. As we shall detail, for large molecules, it is found
possible to reduce the size of the DFT/MRCI configuration space by
orders of magnitude. By applying a perturbative energy correction to
account for the discarded configurations, the errors in the computed
excitation energies can be made essentially negligible ($\sim$1
meV). We denote the combined application of configuration pruning and
perturbative energy corrections to DFT/MRCI wave functions as
p-DFT/MRCI.

The rest of the paper is arranged as follows. In
Section~\ref{sec:dftmrci}, we give a brief overview of the DFT/MRCI
method, focusing on the aspects pertinent to the problem of the
removal of deadwood configurations. Section~\ref{sec:prune_method}
gives the algorithmic details of the p-DFT/MRCI method. In
Section~\ref{sec:results}, we present an analysis of the errors of the
excitation energies furnished by p-DFT/MRCI relative to the original
DFT/MRCI method. In Section~\ref{sec:costs}, we give an analysis of
the computational costs of the p-DFT/MRCI method. Finally, in
Section~\ref{sec:conclusions}, we provide our concluding remarks on
the utility of the proposed methodology and the new research avenues
opened by it.

\section{The DFT/MRCI method}\label{sec:dftmrci}
The electronic Hamiltonian is represented in the DFT/MRCI method in a
basis of CSFs $| \text{w} \omega \rangle$ built from canonical
Kohn-Sham (KS) orbitals. Here, $\text{w}$ denotes a spatial occupation
and $\omega$ a spin-coupling pattern pertaining to the open shells in
$\text{w}$. The total set of CSFs $\Omega = \{ | \text{w} \omega
\rangle \}$ is obtained from the set of single and double excitations
from a small reference space $\Omega_{0} = \{ | \text{w}_{0} \omega
\rangle \}$, where the $\text{w}_{0}$ denote the reference space
configurations. In the following we will refer to $\Omega_{F} = \Omega
\setminus \Omega_{0}$ as the first-order interacting space (FOIS).

The on-diagonal elements of the DFT/MRCI Hamiltonian matrix take the
form of the sum of the exact result and DFT-specific corrections:

\begin{equation}\label{eq:dftmrci_ondiag}
  \left\langle \text{w} \omega \middle| \hat{H}^{DFT} - E_{DFT}
  \middle| \text{w} \omega \right\rangle = \left\langle \text{w}
  \omega \middle| \hat{H} - E_{SCF} \middle| \text{w} \omega
  \right\rangle + \sum_{p} \Delta \text{w}_{p} \left(
  \epsilon_{p}^{KS} - F_{pp} \right) + \Delta E_{x} + \Delta E_{c}.
\end{equation}

\noindent
Here, $\Delta \text{w}_{p} = \text{w}_{p} - \overline{\text{w}}_{p}$
denotes the difference of the occupation of the $p$th spatial orbital
relative to a base, or anchor, occupation $\overline{\text{w}}$,
chosen as the Hartree-Fock occupation. $\epsilon_{p}^{KS}$ and
$F_{pp}$ denote, respectively, the KS orbital energies and on-diagonal
elements of the Fock operator in the KS orbital basis. Finally,
$\Delta E_{x}$ and $\Delta E_{c}$ are Coulomb and exchange
corrections, the exact form of which varies with the different
DFT/MRCI parameterizations\cite{grimme_dft-mrci,
  lyskov_dftmrci_redesign, heil_dftmrci_2017,
  heil_dftmrci_transition_metals}.

The role played by the DFT-specific corrections in
Equation~\ref{eq:dftmrci_ondiag} is to account for the dynamic
electron correlation that would otherwise be described by the coupling
between different CSFs in the MRCI expansion. Thus, the off-diagonal
Hamiltonian matrix elements in question must also be modified to avoid
a double counting of dynamic correlation. This is achieved by
introducing a damping of the off-diagonal elements that is dependent
on the energetic separation of the bra and ket CSFs:

\begin{equation}\label{eq:dftmrci_offdiag}
  \left\langle \text{w} \omega \middle| \hat{H}^{DFT} - E_{DFT}
  \middle| \text{w}' \omega' \right\rangle = \left\langle \text{w}
  \omega \middle| \hat{H} - E_{SCF} \middle| \text{w}' \omega'
  \right\rangle \cdot D(\Delta E_{\text{w}\text{w}'}),
\end{equation}

\noindent
where

\begin{equation}
  \Delta E_{\text{w}\text{w}'} = \frac{1}{n_{\omega}}
  \sum_{\omega}^{n_{\omega}} H_{\text{w}\omega,\text{w}\omega}^{DFT} -
  \frac{1}{n_{\omega'}} \sum_{\omega'}^{n_{\omega'}}
  H_{\text{w}'\omega',\text{w}'\omega'}^{DFT}
\end{equation}

\noindent
denotes the spin coupling-averaged difference between the on-diagonal
matrix elements corresponding to the spatial occupations $\text{w}$
and $\text{w}'$. $D(\Delta E)$ is a some rapidly decaying function,
chosen in practice to be an exponential\cite{grimme_dft-mrci,
  heil_dftmrci_transition_metals} or inverse arctangent
function\cite{lyskov_dftmrci_redesign, heil_dftmrci_2017}.

The effect of damping the off-diagonal Hamiltonian matrix
elements is to decouple CSFs that are energetically well
separated. As well as minimizing the double counting of dynamic
correlation, this also means that CSFs of the reference space,
$\Omega_{0}$, become decoupled from the vast majority of the CSFs
spanning the FOIS, $\Omega_{F}$. This leads to one of the most
important aspects of the DFT/MRCI methodology: the ability to discard
almost all of the FOIS CSFs. In practice, this is achieved using the
following simple energy-based selection criterion. For each FOIS
configuration $\text{w}$, the quantity

\begin{equation}
  d_{\text{w}}= \sum_{p} \Delta \text{w}_{p} \epsilon_{p}^{KS} -
  \delta E_{sel}
\end{equation}

\noindent
is computed, where the parameter $\delta E_{sel}$ is conventionally
chosen as either 0.8 or 1.0 E$_{\text{h}}$\cite{grimme_dft-mrci}. If
$d_{\text{w}}$ is less than the highest reference space eigenvalue of
interest, then all the CSFs generated from the configuration
$\text{w}$ are selected for inclusion, else they are discarded.

\section{The pruned DFT/MRCI method: \MakeLowercase{p}-DFT/MRCI}\label{sec:prune_method}

\subsection{Deadwood in the DFT/MRCI CSF space}
The result of the energy-based selection criterion is a reduction of
the number of CSFs by many orders of magnitude. However, it is found
that a large amount of deadwood CSFs are still present when this is
the sole selection criterion used. By a deadwood CSF we here mean one
that contributes negligibly to the eigenfunctions of interest.

By way of illustration, consider the projection of a given DFT/MRCI
eigenfunction $|\psi_{I}\rangle$ onto the space spanned by the first
$N$ CSFs importance ordered by their absolute coefficient values:

\begin{equation}
  \left| \tilde{\psi}_{I}^{(N)} \right\rangle =
  \sum_{\text{w}\omega}^{N} \left| \text{w} \omega \right\rangle
  \left\langle \text{w} \omega \middle| \psi_{I} \right\rangle
  = \hat{P}_{N} \left| \psi_{I} \right\rangle.
\end{equation}

\begin{figure}
  \begin{center}
    \includegraphics[width=0.5\textwidth,angle=0]{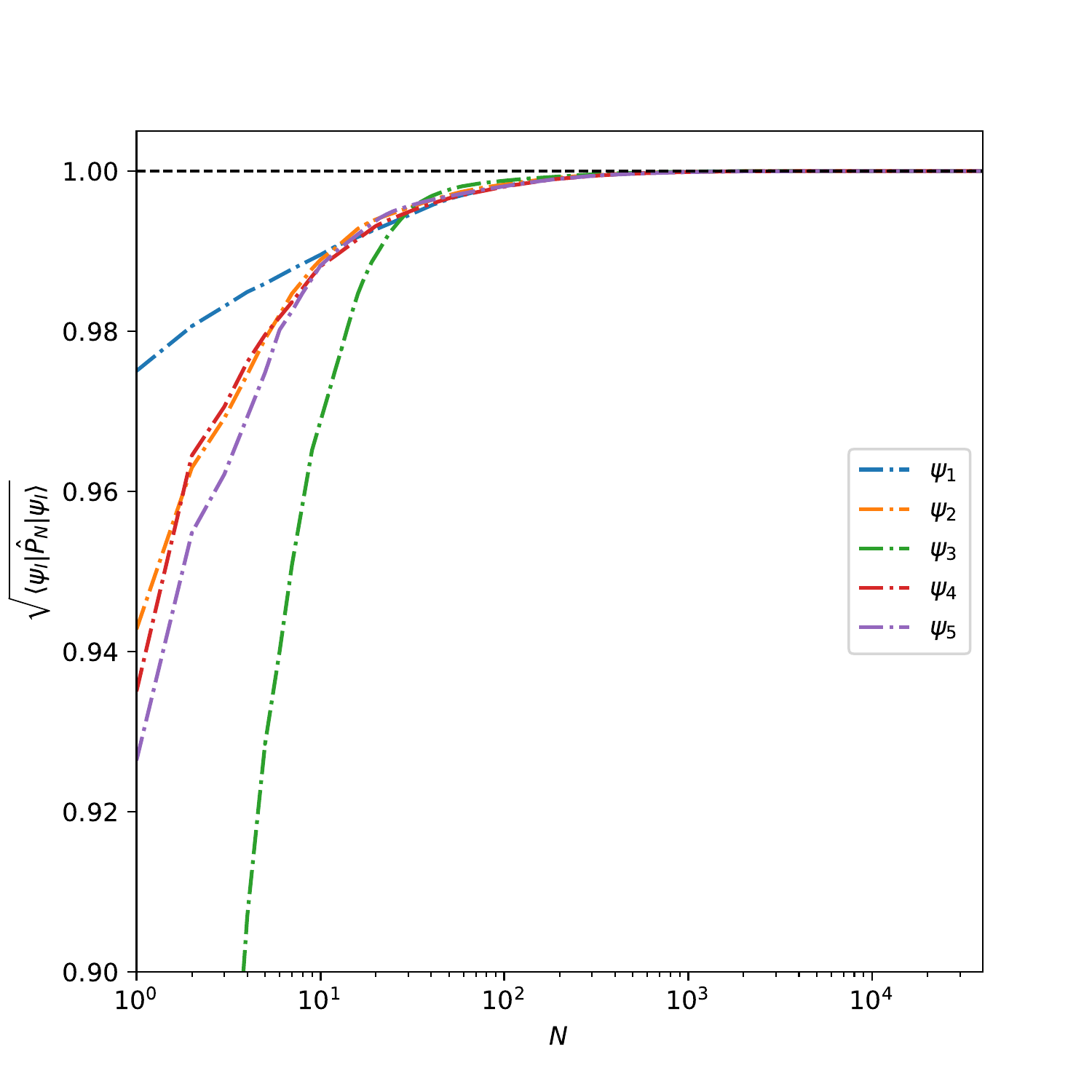}
    \caption{Measure of the ``deadwood'' in DFT/MRCI wave functions:
      the norms of the first five excited state DFT/MRCI/aug-cc-pVDZ
      eigenfunctions of butadiene projected onto the first $N$ highest
      weight CSFs. Wave function norms greater 0.999 are achieved with
      fewer than 200 CSFs per state.}
    \label{fig:butadiene_deadwood}
  \end{center}
\end{figure}

By plotting the norm of $\hat{P}_{N} \left| \psi_{I} \right\rangle$ as
a function of $N$, we may visualize the amount of deadwood present in
the eigenfunction $| \psi_{I} \rangle$. This is shown in
Figure~\ref{fig:butadiene_deadwood} for a DFT/MRCI calculation of the
first five excited states of butadiene computed using the aug-cc-pVDZ
basis. It is clear that with a couple of hundred CSFs, near unit norms
of all five eigenfunctions are attained. This is only a small fraction
of the total of 41681 CSFs that result from the energy-based selection
criterion. Considering that the computational effort of a DFT/MRCI
Hamiltonian build with the number of CSFs has a scaling of
$\mathcal{O}(N^{m})$, $m \sim 1.3-1.5$, we see that significant
computational savings could be realized if the deadwood CSFs could be
identified and removed.

\subsection{Configuration pruning}\label{sec:conf_pruning}
Assume that a reference space diagonalization has been performed,
furnishing reference space eigenpairs $| \psi_{I}^{(0)} \rangle$ and
$E_{I}^{(0)}$, and that the usual energy-based selection criterion has
been used to construct a set $\Omega$ of CSFs. Our aim is to apply a
further pruning of the set $\Omega$, removing those FOIS CSFs that
will not contribute significantly to the final DFT/MRCI
eigenfunctions. To do so, we estimate the contribution to each FOIS
CSF to the final DFT/MRCI eigenfunctions using ENPT2.

In ENPT2, the full CSF space is partitioned into two subspaces, termed
the $P$ and $Q$ spaces. The zeroth-order Hamiltonian, $\hat{H}_{0}$,
is taken to be block diagonal between the $P$ and $Q$ spaces and
diagonal within the $Q$ space. That is

\begin{equation}
    \hat{H}_{0} = \sum_{\text{w} \omega \in P}\sum_{\text{w}' \omega'
      \in P} \Big| \text{w} \omega \Big\rangle \Big\langle \text{w}
    \omega \Big| \hat{H}^{DFT} \Big| \text{w}' \omega' \Big\rangle
    \Big\langle \text{w}' \omega' \Big| + \sum_{\text{w} \omega \in Q}
    \Big| \text{w} \omega \Big\rangle \Big\langle \text{w} \omega
    \Big| \hat{H}^{DFT} \Big| \text{w} \omega \Big\rangle \Big\langle
    \text{w} \omega \Big|,
\end{equation}

\noindent
Within the context of a DFT/MRCI calculation, the natural choice is to
identify the $P$ space with the reference CSFs, and the $Q$ space with
the FOIS CSFs that have survived the energy-based selection
criterion. Then, the ENPT2 approximations to the DFT/MRCI wave
functions read

\begin{equation}
  \left| \psi_{I} \right\rangle \approx \left| \psi_{I}^{(0)}
  \right\rangle + \left| \psi_{I}^{(1)} \right\rangle,
\end{equation}

\noindent
with

\begin{equation}
  \left| \psi_{I}^{(1)} \right\rangle = \sum_{\text{w} \omega \notin
    \Omega_{0}} A_{\text{w} \omega}^{(I)} \left| \text{w} \omega
  \right\rangle,
\end{equation}

\begin{equation}\label{eq:A-vector}
  A_{\text{w} \omega}^{(I)} = \frac{\left\langle \text{w} \omega
    \middle| \hat{H}^{DFT} \middle| \psi_{I}^{(0)}
    \right\rangle}{E_{I}^{(0)} - \left\langle \text{w} \omega \middle|
    \hat{H}^{DFT} - E_{DFT} \middle| \text{w} \omega \right\rangle}.
\end{equation}

The absolute values of the coefficients $A_{\text{w} \omega}^{(I)}$
give a measure of the importance of the FOIS CSF $\text{w} \omega$ to
the $I$th DFT/MRCI eigenfunction. Importantly, these may be determined 
with minimal computational cost using the reference space eigenpairs and the FOIS
CSFs that have passed the energy-based selection criterion.

As DFT/MRCI is most efficiently implemented as a configuration-driven
method, in practice we discard those configurations that do not
generate any CSFs contributing significantly to the ENPT2 first-order
wave function corrections. Our implementation is as follows. The
vectors $\boldsymbol{A}^{(I)}$ of ENPT2 first-order corrections are
computed for all reference space states $| \psi_{I}^{(0)}
\rangle$. For each state, the smallest subset $S_{I}$ of FOIS
configurations satisfying

\begin{equation}\label{eq:pruning}
  \sum_{\text{w} \in S_{I}} \sum_{\omega} \big|
  A_{\text{w}\omega}^{(I)} \big|^{2} \ge \alpha_{p} \big|
  \boldsymbol{A}^{(I)} \big|^{2}
\end{equation}

\noindent
are determined, where $\alpha_{p} < 1$. The final subset of FOIS
configurations chosen for inclusion in the p-DFT/MRCI calculation is
then taken as the union of the subsets $S_{I}$ for each state. As the
vast majority of FOIS configurations have already been removed by the
energy-based selection criterion, this additional pruning step is
extremely cheap. Furthermore, the accuracy of the p-DFT/MRCI
eigenpairs (relative to the unpruned ones) is systematically
controlled by the single parameter $\alpha_{p}$.

It remains to note that, in practice, if the $n$ lowest energy
DFT/MRCI states are required, they will not generally correspond to
the first $n$ reference space states. As such, we include a small
number $n_{extra} \approx$ 10 of extra reference space states, and
pick the $n$ roots with the lowest ENPT2-corrected energies for use in
the configuration pruning.

\subsection{ENPT2 energy corrections}\label{sec:enpt2_correction}
The configuration pruning algorithm presented in
Section~\ref{sec:conf_pruning} is controlled by the pruning threshold
$\alpha_{p}$ that enters into Equation~\ref{eq:pruning}. The smaller
the value of $\alpha_{p}$, the greater the proportion of FOIS CSFs are
discarded, but to the increasing detriment of the resulting p-DFT/MRCI
energies. In order to use as small a pruning threshold as possible,
the contributions of the discarded CSFs to the DFT/MRCI energies can
be accounted for perturbatively. To do so, the ENPT2 energy
corrections

\begin{equation}\label{eq:enpt2_corr}
  E_{I,\text{w}\omega}^{(2)} = \frac{\left| \left\langle \text{w}
    \omega \middle| \hat{H}^{DFT} \middle| \psi_{I}^{(0)}
    \right\rangle \right|^2}{E_{I}^{(0)} - \left\langle \text{w}
    \omega \middle| \hat{H}^{DFT} - E_{DFT} \middle| \text{w} \omega
    \right\rangle}
\end{equation}

\noindent
are computed for the discarded CSFs. These corrections are then added
to the energies computed using the pruned CSF basis. In practice, this
allows for a significantly greater proportion of FOIS CSFs to be
discarded while maintaining the accuracy of the resultant
energies. Importantly, these energy corrections are essentially
``free'' with the evaluation of the first-order wave function
corrections (Equation~\ref{eq:A-vector}). From here on out, we will
use the term p-DFT/MRCI to refer to the combination of configuration
pruning \textit{and} the application of the ENPT2 energy corrections
(Equation~\ref{eq:enpt2_corr}).

\subsection{Automated selection of the reference space}\label{sec:autoras}
In a standard (unpruned) DFT/MRCI calculation the initial choice of
reference space configurations is not particularly critical to the
quality of the resulting DFT/MRCI wave functions. The reason for this
is that a DFT/MRCI calculation is typically performed in the following
iterative fashion. First, a guess reference space is selected, and the
DFT/MRCI wave functions are computed using this initial space. Next,
the reference space is updated to contain the dominant configurations
present in the DFT/MRCI wave functions, and the calculation is
repeated. This reference space refinement is continued until
convergence of the DFT/MRCI wave functions is attained, a process
usually requiring only a couple of iterations.

However, the use of Epstein-Nesbet perturbation theory in p-DFT/MRCI
is predicated on the reference space states $| \psi_{I}^{(0)} \rangle$
being good zeroth-order descriptions of the states of interest. As
such, the configuration pruning and ENPT2 energy correction algorithms
can yield poor results if the initial reference space is not chosen
reasonably. For large numbers of states, a good choice of the initial
reference space may not be obvious, and it is desirable to automate
its generation. In order to arrive at a robust, fully automated
algorithm, we first note that, from experience, the use of a
restricted active space CI (RASCI)\cite{olsen1988} set of
configurations with only the RAS1 (characterized by a maximum number
of holes) and RAS3 (defined by a maximum number of electrons)
subspaces occupied is almost always a good choice for the initial
reference space. The problem then shifts to the selection of the RAS1
and RAS3 MOs.

To automatically select the RAS1 and RAS3 MOs, we have explored the
use of a computationally cheap, preliminary combined DFT and CIS (DFT/CIS)
calculation\cite{grimme_dftcis}, selecting the particle/hole MOs that
appear in the dominant DFT/CIS configurations. In practice we have
found this to be a robust route to the generation of initial reference
spaces that yield good quality zeroth-order wave functions. However,
for large systems, this preliminary DFT/CIS calculation can become
more expensive than the pruned DFT/MRCI calculation itself. To
overcome this, an aggressively loose integral pre-screening is
introduced into the DFT/CIS calculation, making use of the
Cauchy-Schwarz inequality

\begin{equation}
  \left| (pq|rs) \right| \le \sqrt{(pq|pq)} \sqrt{(rs|rs)} = G_{pq}
  G_{rs}.
\end{equation}

\noindent
In evaluating the DFT/CIS equations, off-diagonal Hamiltonian matrix
elements $H_{mn}$ with bra and ket configurations corresponding to
excitation from MOs $i \rightarrow a$ and $j \rightarrow b$ are
skipped if

\begin{equation}
  \frac{1}{2} G_{ij} G_{ab} < \tau \hspace{0.5cm}
  \text{and} \hspace{0.5cm} 2 G_{ia} G_{jb} < \tau.
\end{equation}

\noindent
In practice, using a large integral screening threshold $\tau =
\mathcal{O}(10^{-2})$ is found to furnish a rather severe, but
uniform, degradation of the quality of the resulting DFT/CIS energies.
As such, the particle/hole MOs of the dominant configurations may
still be reliably identified, but at a much reduced cost.

\subsection{Implementation and computational details}\label{sec:implementation}
All p-DFT/MRCI calculations were performed using a newly developed
software package tailored for the calculation of excited electronic
states using CI with generalized reference spaces\cite{graci_github}.
The required KS MOs, energies and integrals were computed using the
PySCF package\cite{sun_2018, sun_2020}. The density fitting
approximation\cite{whitten_1973, feyereisen_1993, vahtras_1993} was
used in the evaluation of the two-electron integrals. Following the
original DFT/MRCI implementation by Grimme and
Waletzke\cite{grimme_dft-mrci}, ideas from the works of Segal
\textit{et al.}\cite{wetmore_1975, segal_1978} and Engels and
Hanrath\cite{hanrath_1997} were used in the evaluation of the DFT/MRCI
Hamiltonian matrix elements. Additionally, aspects of the
bitstring-based algorithms of Scemama, Giner and Garniron were adapted
for the computation of spin-coupling coefficients and the
identification of non-zero Hamiltonian matrix elements in a CSF
basis\cite{garniron_thesis, scemama_slater_condon}.

The automatic initial reference space generation algorithm described
in Section~\ref{sec:autoras} is based on a preliminary DFT/CIS
calculation. The original DFT/CIS Hamiltonian was parameterised for
use with the B3LYP functional\cite{grimme_dftcis}. However, all
current DFT/MRCI implementations are parameterised for use with the
BHLYP functional. Ideally, then, the parameters of the DFT/CIS would
be reoptimized for use with the same functional, but this is somewhat
beyond the scope of this preliminary work. As such, in a first step,
the parameters of Grimme's DFT/CIS Hamiltonian were used, with the
Coulomb scaling parameter ($c_{1}$ in the notation of Reference
\citenum{grimme_dftcis}) manually adjusted to yield acceptable
excitation energies for a small number of molecules. In all
calculations described here, a value of $c_{1} =$ 0.596 was used.

To assess the efficacy of the p-DFT/MRCI methodology, vertical
excitation energies were computed for the molecules in Thiel's test
set of 28 small-to-medium sized organic
molecules\cite{schreiber_2008}. For each molecule, four singly excited
states of each symmetry were computed, resulting in a total of 472
excitation energies, including those of both valence and Rydberg
character. In all cases, the original DFT/MRCI Hamiltonian
parameterization of Grimme and Waletzke\cite{grimme_dft-mrci} was
used, although we expect the conclusions drawn to apply similarly to
the re-designed Hamiltonians of Marian \textit{et
  al.}\cite{lyskov_dftmrci_redesign, heil_dftmrci_2017,
  heil_dftmrci_transition_metals} All calculations were performed
using the aug-cc-pVDZ basis\cite{dunning_1989} and aug-cc-pVDZ-jkfit
auxiliary basis.

\section{Results}\label{sec:results}

\begin{figure}
  \begin{center}
    \includegraphics[width=\textwidth,angle=0]{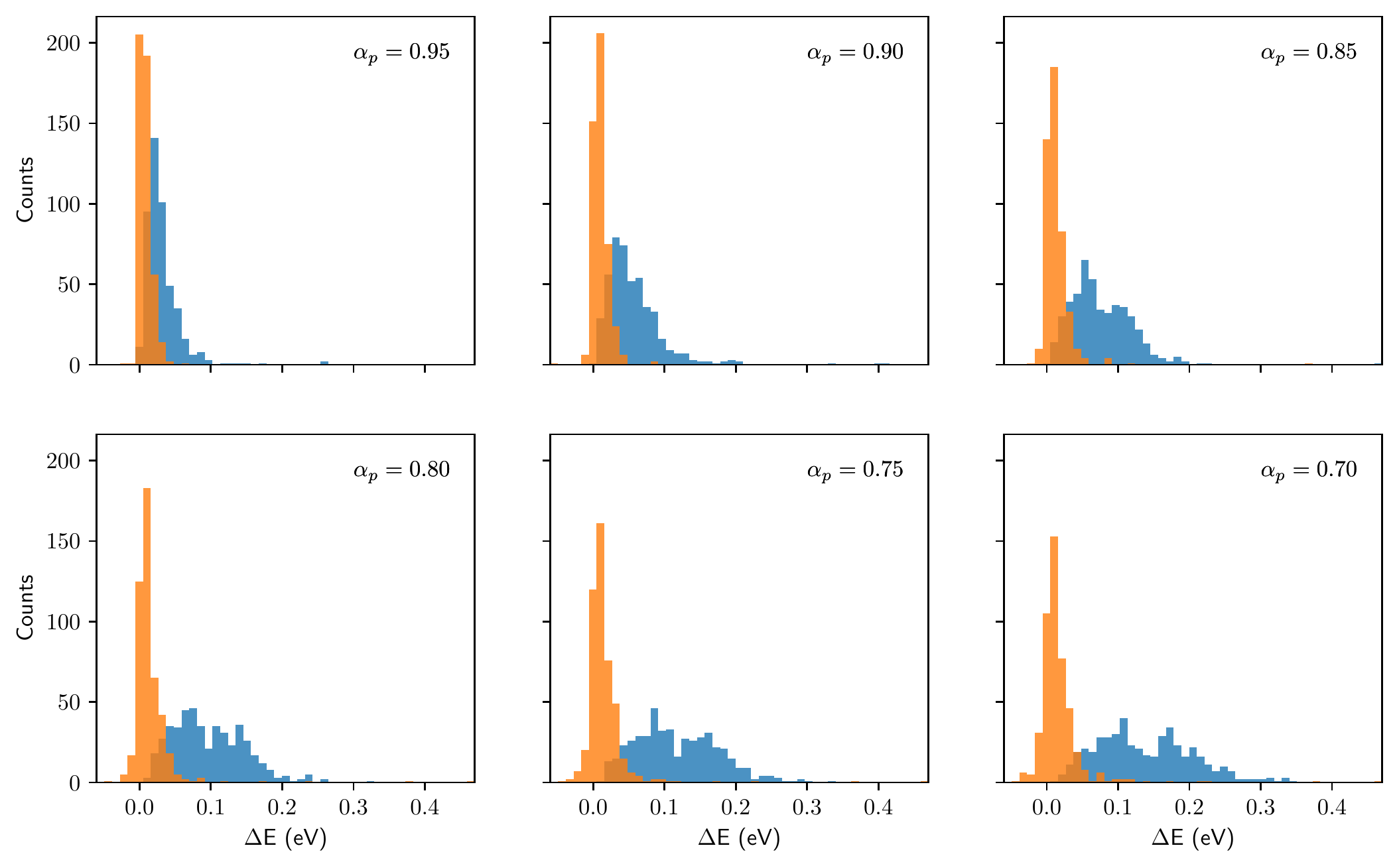}
    \caption{Excitation energy errors as a function of the pruning
      threshold $\alpha_{p}$ relative to the unpruned values. Orange:
      ENPT2 energy-corrected values. Blue: uncorrected values.}
    \label{fig:deltae_comp}
  \end{center}
\end{figure}

\subsection{Error analysis}\label{sec:error}
We first consider the errors introduced as a function of the pruning
threshold $\alpha_{p}$. To do so, all 472 excitation energies in the
test set were computed using a fixed energy-based selection threshold
of $\delta E_{sel}=1.0$ E$_{\text{h}}$ and $\alpha_{p}$ set to values
between 1.0 (unpruned) and 0.70. Shown in Figure~\ref{fig:deltae_comp}
are the differences, $\Delta E$, in the computed excitation energies
relative to the unpruned values as a function of $\alpha_{p}$. For
comparison, the differences $\Delta E$ for both the uncorrected (blue)
and ENPT2 energy-corrected (orange) values are shown.

With the application of the ENPT2 energy correction, for the two
tightest thresholds ($\alpha_{p}=$ 0.95 and $\alpha_{p}=$ 0.90), the
errors introduced by the configuration pruning procedure are extremely
small, with root mean square deviations (RMSDs) not exceeding 0.015
eV. Decreasing the value of $\alpha_{p}$ steadily increases the RMSD
of the computed ENPT2-corrected excitation energies. However, even
with the rather loose pruning threshold of $\alpha_{p}=$ 0.7, an RMSD
of only 0.041 eV is attained. At this level of pruning, FOIS CSFs that
contribute non-negligibly to the DFT/MRCI wave functions are being
discarded. However, the use of ENPT2 energy corrections (see
Section~\ref{sec:enpt2_correction}) is able to compensate for
this.

\begin{figure}
  \begin{center}
    \includegraphics[width=\textwidth,angle=0]{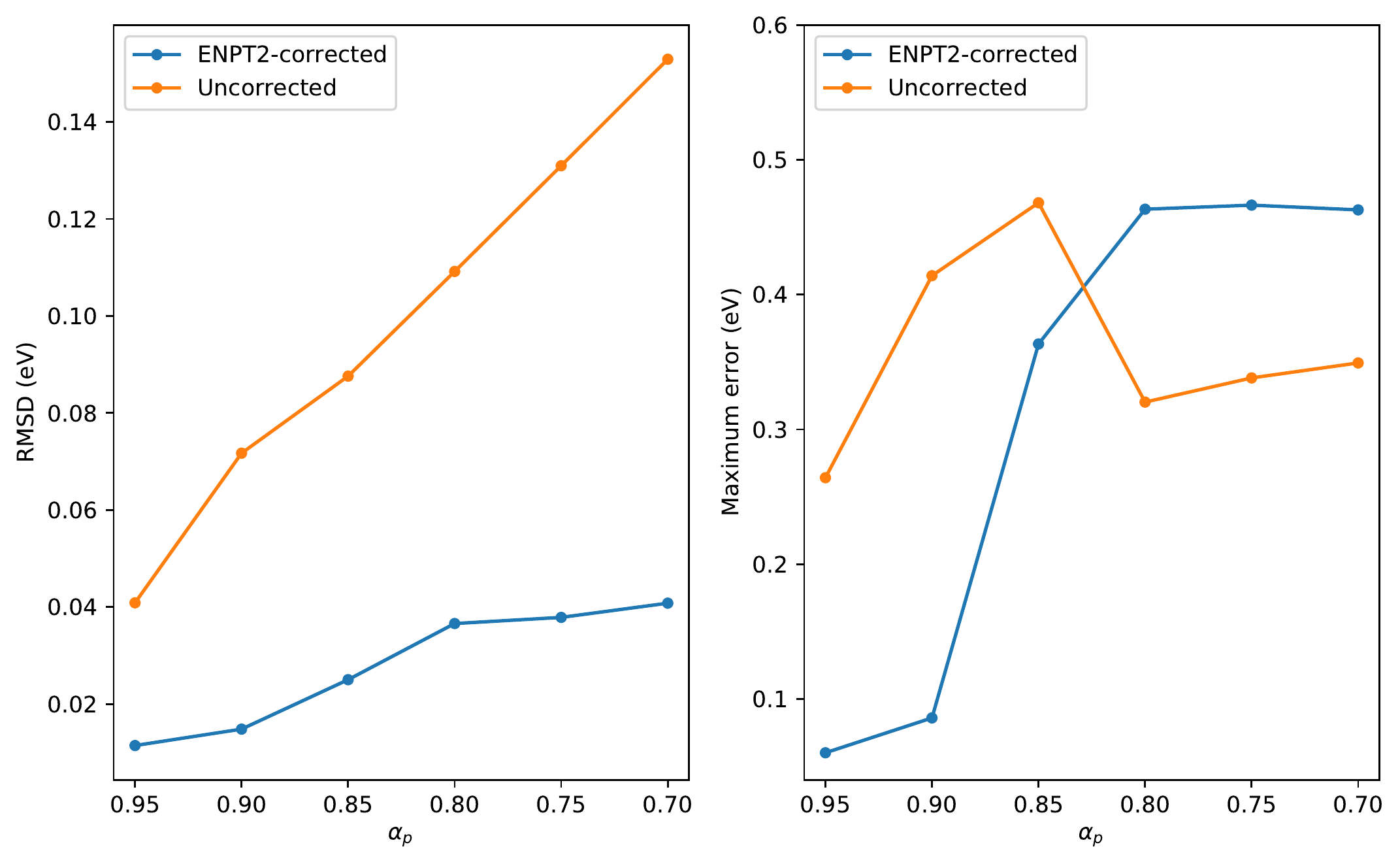}
    \caption{Comparison of RMSDs and maximum errors of excitation
      energies as a function of the pruning threshold $\alpha_{p}$,
      both with and without the ENPT2 energy correction applied. All
      values correspond to an energy-based selection threshold of
      $\delta E_{sel}=1.0$ E$_{\text{h}}$.}
    \label{fig:error}
  \end{center}
\end{figure}

The effect of the ENPT2 energy corrections is clearly illustrated by
the errors for the uncorrected excitation energies (shown in orange in
Figure~\ref{fig:deltae_comp}). Here, a much more severe degradation in
the quality of the excitation energies with decreasing pruning
threshold value is observed, highlighting the importance of the ENPT2
energy corrections. For reference, we also show in
Figure~\ref{fig:error} a comparison of the RMSDs of the excitation
energies as a function of $\alpha_{p}$ both with and without the
application of the ENPT2 energy corrections. As may be expected, the
RMSDs for the non-corrected excitation energies scale approximately
linearly with $\alpha_{p}$. Application of the ENPT2 energy
corrections clearly breaks this relationship, particularly for smaller
values of $\alpha_{p}$.

Also shown in Figure~\ref{fig:error} are the maximum errors as a
function of the pruning threshold. Here, the situation appears to be
worse, with maximum errors of greater than 0.3 eV being seen for all
values of $\alpha_{p}<$ 0.90, both with and without the application of
the ENPT2 energy corrections. In fact, the maximum errors for values
of $\alpha_{p}<$ 0.85 are larger for the ENPT2 energy corrected values
than for the uncorrected ones. In most cases the largest errors in the
ENPT2 energy-corrected values arise due to the extremal roots computed
in the pruned and unpruned calculations corresponding to different
states. This can usually be ameliorated by computing a handful of
extra roots. However, in order to avoid this problem, it seems seems
advisable to not use a pruning threshold $\alpha_{p}$ significantly
below 0.90.

\subsection{Reduction in size of the CSF basis }
Having established the generally favorable errors introduced by the
combination of configuration pruning and ENPT2 energy corrections, we
now turn our attention to the computational gains afforded by the
p-DFT/MRCI method. To do so, we consider the reduction of the CSF
basis size as a function of the pruning threshold. Here, we consider
only values of $\alpha_{p}$ = 0.95 and 0.90, which, as discussed in in
Section \ref{sec:error}, are found to consistently yield small maximum
errors.

As a measure of the speedups resulting from the configuration pruning
algorithm, we consider the CSF basis compression factor

\begin{equation}
  \kappa(\alpha_{p}) =
  \frac{N_{CSF}(\alpha_{p}=1)}{N_{CSF}(\alpha_{p})}.
\end{equation}

\noindent
That is, the ratio of the number of CSFs without and with the
application of configuration pruning as a function of the pruning
threshold $\alpha_{p}$. We note that the cost of a single DFT/MRCI
Hamiltonian build scales empirically as $\mathcal{O}(N_{CSF}^{m})$, $m
\sim 1.3-1.5$. Thus, although the individual configuration selection
inherent to the DFT/MRCI method precludes an exact scaling
relationship, we may expect speedups of the order of
$\kappa^{m}(\alpha_{p})$, $m \sim 1.3-1.5$ to be afforded by the
p-DFT/MRCI method.

\begin{figure}
  \begin{center}
    \includegraphics[width=0.5\textwidth,angle=0]{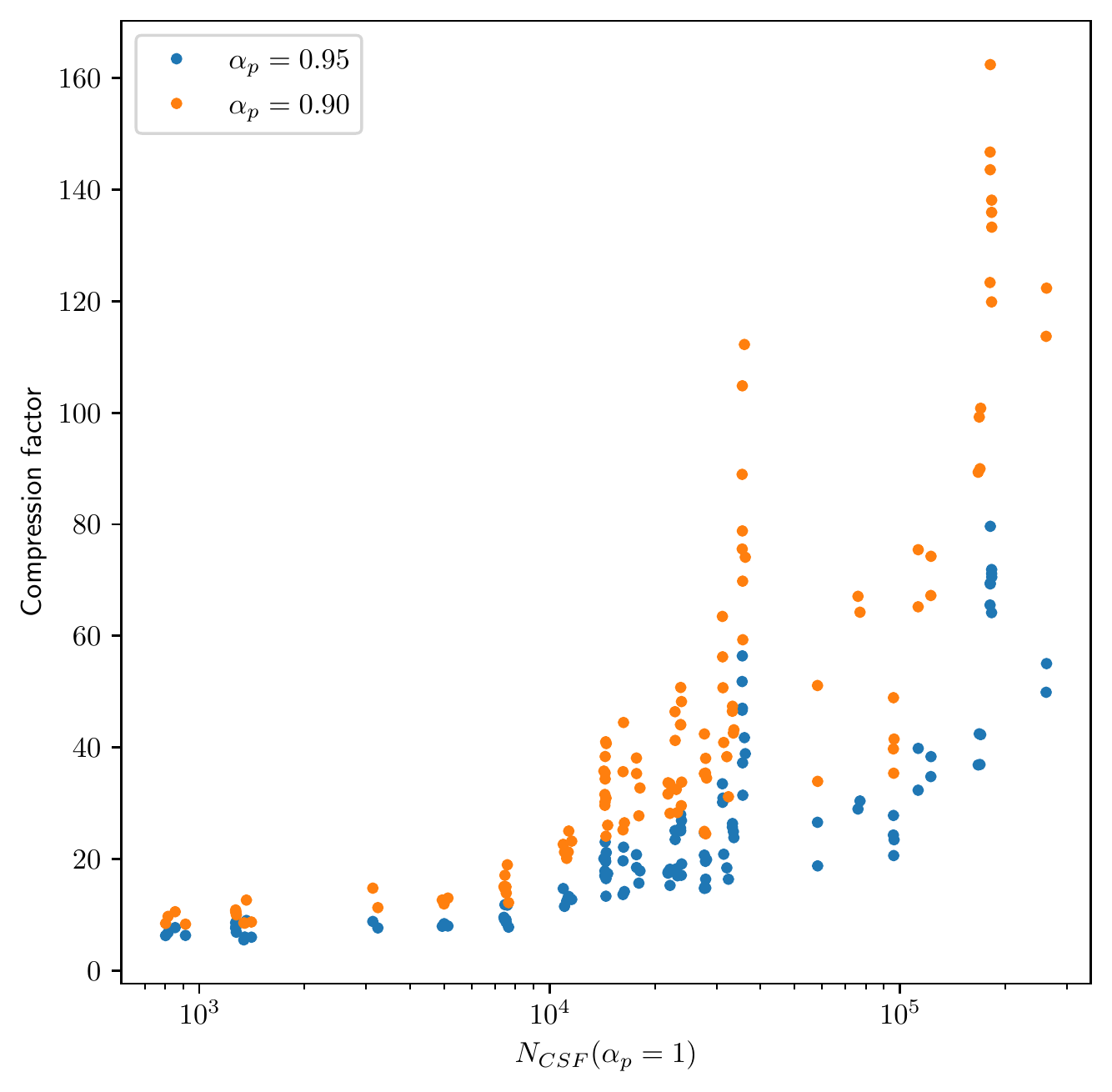}
    \caption{CSF basis compression factors as a function of the
      unpruned number of CSFs, $N_{CSF}(\alpha_{p}=1)$, for pruning
      thresholds $\alpha_{p}$ of 0.95 and 0.90. All values correspond
      to an energy-based selection threshold of $\delta E_{sel}=1.0$
      E$_{\text{h}}$.}
    \label{fig:compression}
  \end{center}
\end{figure}

Shown in Figure~\ref{fig:compression} are the CSF basis compression
factors as a function of the number of CSFs in the unpruned basis. For
both $\alpha_{p}=$ 0.95 and 0.90, there is a clear trend towards
higher compression factors with increasing size of the unpruned CSF
basis. That is, the computational savings introduced by configuration
pruning are greatest for larger systems. For $\alpha_{p}=$ 0.95,
maximal compression factors in the range 60-80 are found. For
$\alpha_{p}=$ 0.90, this increases to around 100-160. Assuming a lower
bound of $\mathcal{O}(N_{CSF}^{1.3})$ for the scaling of the cost of a
DFT/MRCI Hamiltonian build, this translates into conservative speedup
estimates of the order of 100-1000$\times$. It is important to note
that the largest molecule present in the test set (naphthalene) is
still relatively small, and that, from the trends observed, we
anticipate even greater gains upon increasing molecular size.

Finally, we note that the average CSF basis compression factor for
$\alpha_{p}=$ 0.90 is 1.8 times higher than for $\alpha_{p}=$
0.95. Considering the negligible increase in the error introduced by
this change in pruning threshold, we recommend using a value of
$\alpha_{p}=$ 0.90 for an optimal balance of accuracy and computation
effort.

\subsection{Energy-based configuration selection}
Separate from the configuration pruning algorithm presented here,
there does exist another mechanism for reducing the size of the
DFT/MRCI CSF basis. Namely, the reduction of the energy-based
configuration selection threshold $\delta E_{sel}$ (see
Section~\ref{sec:dftmrci}). Indeed, all existing DFT/MRCI Hamiltonians
have been separately parameterised for values of both $\delta
E_{sel}=$ 1.0 E$_{\text{h}}$ and $\delta E_{sel}=$ 0.8
E$_{\text{h}}$. When the smaller of the two values, a significant
reduction of the CSF basis does result. Thus, it is desirable to
consider the errors and CSF basis reductions introduced by using
$\delta E_{sel}=$ 0.8 E$_{\text{h}}$, and to compare these with those
afforded by the configuration pruning algorithm.

\begin{figure}
  \begin{center}
    \includegraphics[width=0.45\textwidth,angle=0]{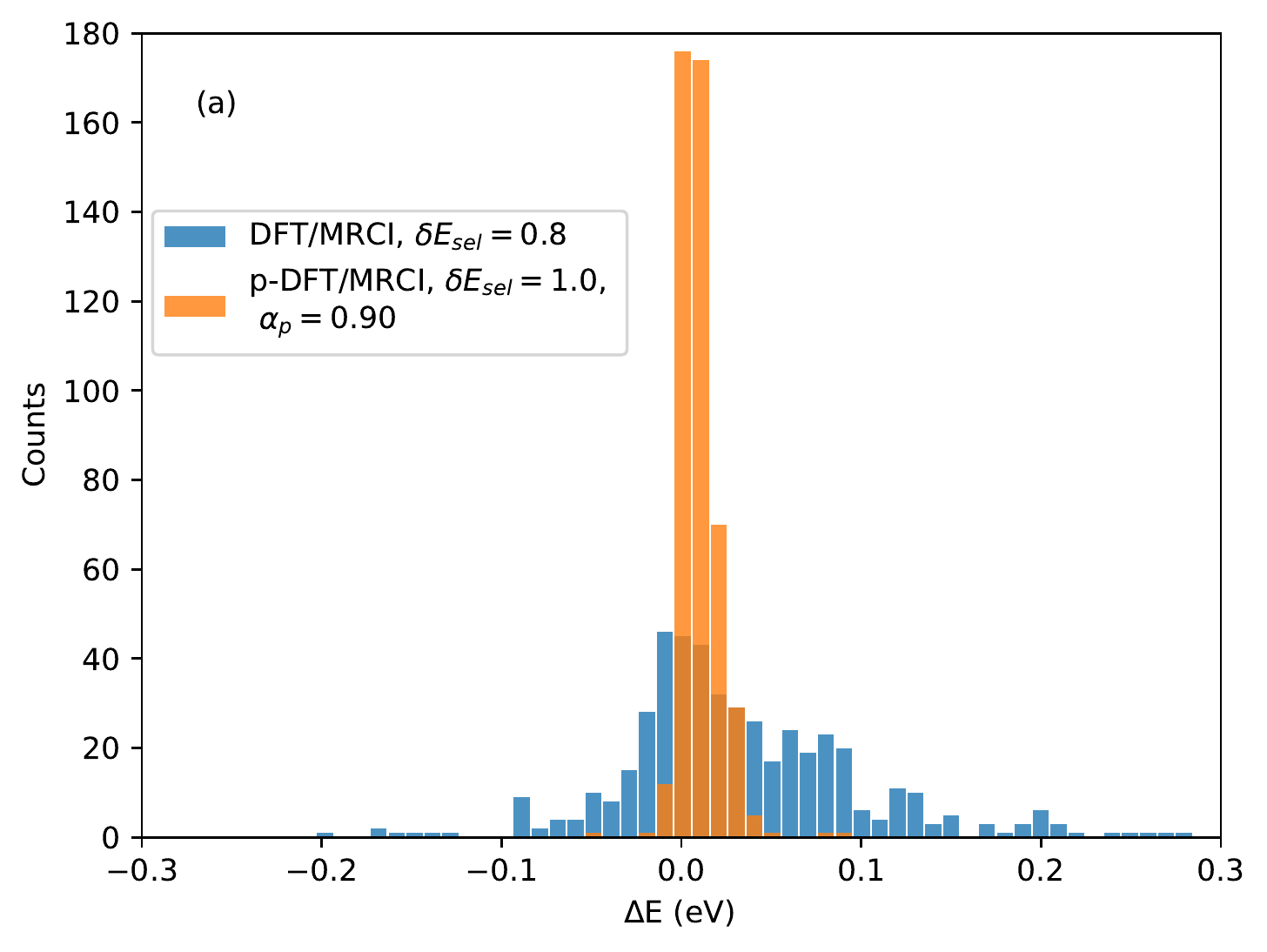}
    \includegraphics[width=0.44\textwidth,angle=0]{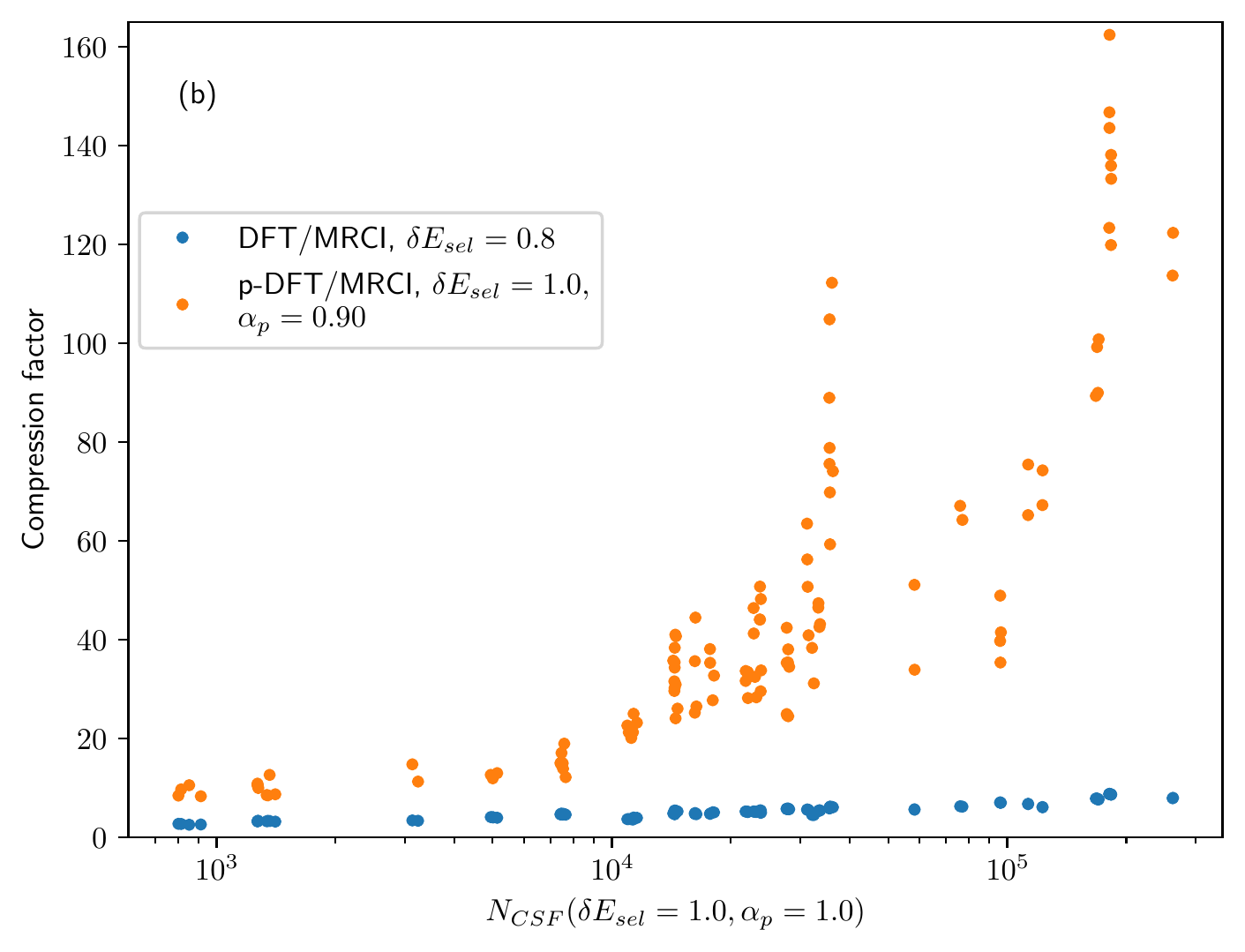}
    \caption{Comparison of: (a) excitation energy errors, and (b) CSF
      basis compression factors introduced decreasing energy-based
      selection threshold $\delta E_{sel}$ (blue) and using
      configuration pruning (orange). All values are taken relative to
      the unpruned ($\alpha_{p}=1.0$) calculations using an
      energy-based selection threshold of $\delta E_{\text{sel}}=1.0$
      E$_{\text{h}}$.}
    \label{fig:short_results}
  \end{center}
\end{figure}

Shown in Figure~\ref{fig:short_results} are both the errors in the
excitation energies and the CSF basis compression factors introduced
upon reduced the energy-based configuration selection threshold,
$\delta E_{sel}$, from 1.0 to 0.8 E$_{\text{h}}$. For reference, the
corresponding values resulting from p-DFT/MRCI calculations using
selection thresholds $\delta E_{sel}=1.0$ E$_{\text{h}}$ and
$\alpha_{p}=0.90$ are shown alongside. Reducing $\delta E_{sel}$ from
1.0 to 0.8 E$_{\text{h}}$ results in excitation energy errors with an
RMSD 0.074 eV is found. This is to be compared to values of less than
0.015 eV for the p-DFT/MRCI results with $\alpha_{p} \le$
0.90. Moreover, for the test set considered, there exist no CSF basis
compression factors greater than 10. This is to be compared with a
maximum value of $\sim$160 for the p-DFT/MRCI results. We thus
conclude that the configuration pruning algorithm is to be favored
over decreasing the energy-based configuration selection threshold,
$\delta E_{sel}$, both in terms of accuracy and computational effort.

\subsection{Computational costs}\label{sec:costs}
\subsubsection{CPU times}

\begin{figure}
  \begin{center}
    \includegraphics[width=0.5\textwidth,angle=0]{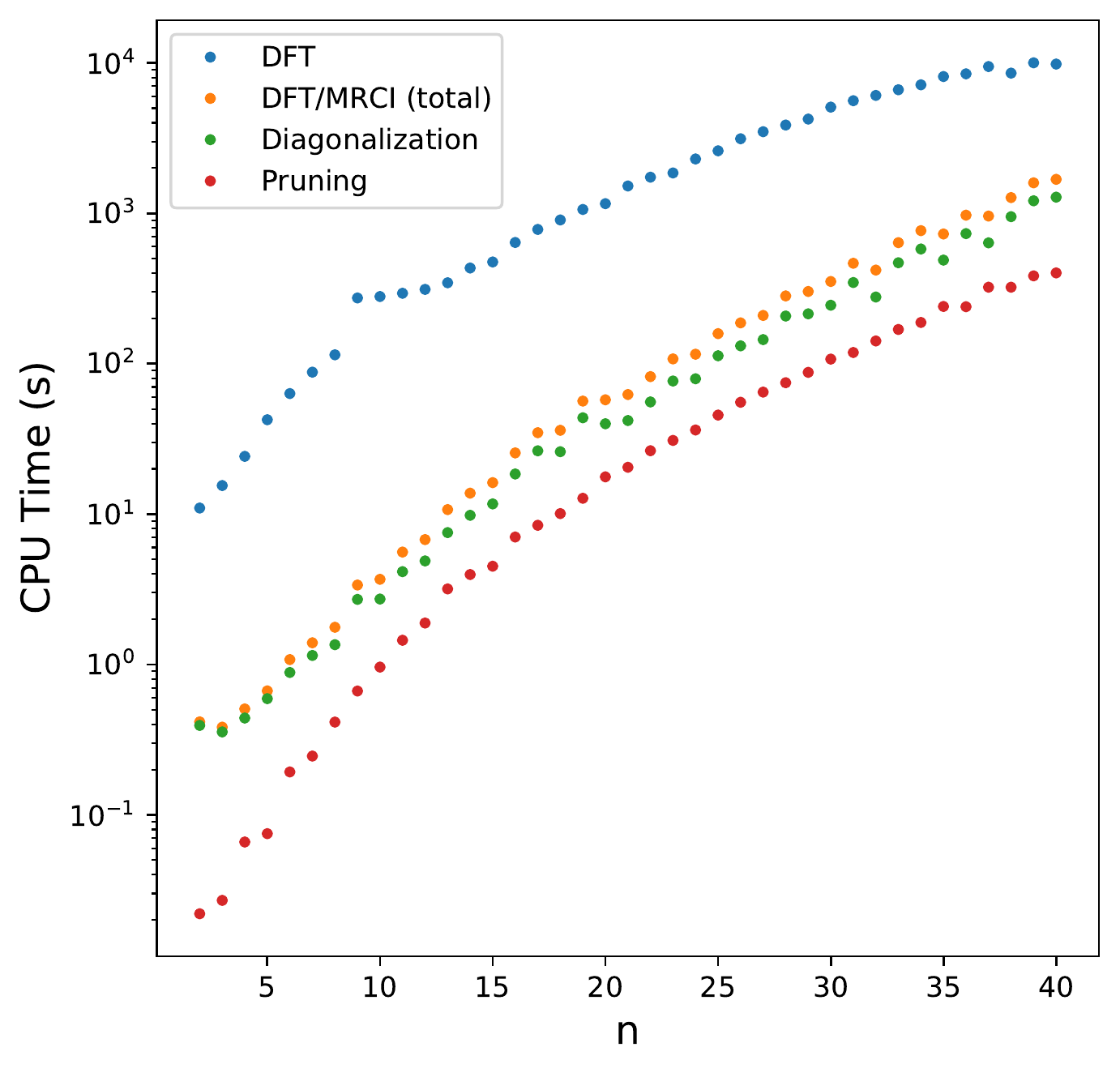}
    \caption{Timings for the main steps involved in a pruned DFT/MRCI
      calculation of the first excited state of each symmetry for a
      series of linear alkanes C$_{n}$H$_{2n+2}$. The def2-SVP basis
      set was used along with a pruning threshold $\alpha_{p}=$ 0.90
      and an energy-based selection threshold of $\delta E_{sel}=$ 1.0
      E$_{\text{h}}$. All calculations were performed on a single
      Intel Xeon E5-2660 v4 CPU core.}
    \label{fig:alkanes}
  \end{center}
\end{figure}

We end by considering the computational costs p-DFT/MRCI calculations
for a representative example: the series of linear alkanes
C$_{n}$H$_{n+2}$ for $n$ up to 40. In these calculations, the first
excited state of each symmetry was computed (a total of five roots,
including the ground state). The def2-SVP basis and def2-SVP-jkfit
auxiliary basis were used along with a pruning threshold of
$\alpha_{p}=$ 0.90. Shown in Figure~\ref{fig:alkanes} are the
calculation timings obtained using single Intel Xeon E5-2660 v4 CPU
core. Here, the timings are broken down into the three main steps: the
preceding DFT calculation, and the configuration pruning and the
DFT/MRCI Hamiltonian diagonalization steps. Although the absolute
timings are hardware- and implementation dependent, some useful
conclusions may still be drawn. First, it is evident that the pruning
step introduces very little overhead, with the total p-DFT/MRCI
calculation being dominated by the Hamiltonian diagonalization
step. Second, for all members of the series C$_{n}$H$_{n+2}$ up to
$n=$ 40, the preceding DFT calculation (performed using the PySCF
package\cite{sun_2018, sun_2020}) is significantly more expensive than
the subsequent p-DFT/MRCI calculation. The ratio of the two timings
does decrease with increasing system size. However, even for the
largest system size considered (C$_{40}$H$_{82}$, 322 electrons, 970
MOs), the p-DFT/MRCI calculation is still around six times quicker
that the prerequisite DFT calculation. Important to note here is that
this performance was attained using a relatively inefficient pilot
implementation of the DFT/MRCI method, and this ratio will only
increase with code optimization and algorithmic development.

\subsubsection{Memory requirements}
Although the use of configuration pruning helps to alleviate the
Hamiltonian diagonalization bottleneck, the p-DFT/MRCI method is
currently practically limited to around 1500-2000 correlated MOs. The
reason for this is a steeply increasing memory cost associated with
the storage of two-electron integrals.

Common to all selected CI methods, DFT/MRCI suffers from the property
that $\sigma$-vector (i.e., Hamiltonian matrix-vector product)
calculations cannot efficiently be made integral-driven. Instead, a
configuration-driven algorithm is used, in which `arbitrary' sets of
two-electron integrals are required to be accessed on an
element-by-element basis. In turn, this necessitates the in-core
storage of the two-electron integrals, which in practice are
decomposed using density fitting (DF) to reduce storage costs. Even
with the use of DF, however, the memory cost still scales as
$\mathcal{O}(N_{MO}^{3})$, resulting in tremendous costs for $N_{MO}
\gtrsim$2000. By way of example, extending the linear alkanes example
to $n$=100 (C$_{100}$H$_{202}$) would require 483 GB RAM for the
storage of the two-electron integrals.

Clearly, in order to extend the p-DFT/MRCI method to larger systems, a
different route must be perused. One potential strategy would be to
use a different decomposition of the two-electron integral tensor. A
promising candidate is the tensor hypercontraction (THC) method of
Mart\'{i}nez \textit{et al.}\cite{hohenstein_thc1, parrish_thc2,
  hohenstein_thc3}. Using THC in place of DF, the integral storage
cost would be decrease to $\mathcal{O}(N_{MO}^{2})$, significantly
increasing the size of system amenable to application of the
p-DFT/MRCI method.

\section{Conclusions}\label{sec:conclusions}
We have presented an approach for the removal of deadwood
configurations from DFT/MRCI calculations. The proposed algorithm,
termed p-DFT/MRCI, is based on a computationally cheap pruning step,
utilizing Epstein-Nesbet perturbation theory to estimate the
contribution of each individual configuration to the eigenstates of
interest. By combining this pruning scheme with a simple,
state-specific energy correction to account for the pruned
configurations, a reduction of the CSF basis size by up to two orders
of magnitude can be achieved while introducing negligible errors.

A pre-requisite of the configuration pruning algorithm is that an
initial reference space with good support of the desired eigenstates
be used. To address this, an automated initial reference space
generation algorithm was devised and implemented, based on an
approximate preliminary DFT/CIS calculation. The result is a black box
method controlled by a single pruning threshold parameter,
$\alpha_{p}$.

From the test set of small-to-medium sized molecules considered here,
it is apparent that the proportion of configurations removed by the
pruning scheme increases significantly with the system
size. Accordingly, we expect that the already impressive CSF basis
reduction factors reported here will only increase when larger
molecules are considered.

It remains to note that, although the use configuration pruning
alleviates the Hamiltonian diagonalization bottleneck to a significant
degree, the p-DFT/MRCI method is still limited by an integral storage
cost scaling as $\mathcal{O}(N_{MO}^{3})$ in its current DF-based
implementation. This should be alleviated by adopting a lower scaling
decomposition of the two-electron integral tensor. The use of THC
decomposition appears highly promising in this regard and will be the
subject of future work, laying the way for the application of
p-DFT/MRCI to systems of unprecedented size.



\providecommand{\latin}[1]{#1}
\makeatletter
\providecommand{\doi}
  {\begingroup\let\do\@makeother\dospecials
  \catcode`\{=1 \catcode`\}=2 \doi@aux}
\providecommand{\doi@aux}[1]{\endgroup\texttt{#1}}
\makeatother
\providecommand*\mcitethebibliography{\thebibliography}
\csname @ifundefined\endcsname{endmcitethebibliography}
  {\let\endmcitethebibliography\endthebibliography}{}

\end{document}